\documentclass[sigconf]{acmart}
\AtBeginDocument{%
  }

\copyrightyear{2026}
\acmYear{2026}
\setcopyright{cc}
\setcctype{by}

\acmConference[MM '26]
{Proceedings of the 34th ACM International Conference on Multimedia}
{November 10--14, 2026}
{Rio de Janeiro, Brazil.}

\acmBooktitle{Proceedings of the 34th ACM International Conference on Multimedia
(MM '26), November 10--14, 2026, Rio de Janeiro, Brazil}

\acmISBN{979-8-4007-2213-4/2026/11}
\acmDOI{10.1145/3767308.3834906}
\usepackage{multirow}
\usepackage{colortbl}
\usepackage{pifont}

\definecolor{best}{RGB}{255, 200, 200}
\definecolor{second}{RGB}{218, 232, 252}

\begin{document}

\title{SAMOT: State-Aware Step Modulation and Optimal Transport Matching for Audio-Visual Instance Segmentation}

\author{Kai Peng}
\authornote{Kai Peng and Yunzhe Shen contributed equally to this work.}
\affiliation{%
  \institution{Dalian University of Technology}
  \city{Dalian}
  \country{China}}
\email{happypk@mail.dlut.edu.cn}

\author{Yunzhe Shen}
\authornotemark[1]
\affiliation{%
  \institution{Dalian University of Technology}
  \city{Dalian}
  \country{China}}
\email{1079006460@mail.dlut.edu.cn}

\author{Miao Zhang}
\correspondingauthor
\affiliation{%
  \institution{Dalian University of Technology}
  \city{Dalian}
  \country{China}}
\email{miaozhang@dlut.edu.cn}

\author{Leiye Liu}
\affiliation{%
  \institution{Dalian University of Technology}
  \city{Dalian}
  \country{China}}
\email{leiyeliu@mail.dlut.edu.cn}

\author{Wei Ji}
\affiliation{%
  \institution{Yale University}
  \city{New Haven}
  \country{United States}}
\email{wei.ji@yale.edu}

\author{Jingjing Li}
\affiliation{%
  \institution{Carnegie Mellon University}
  \city{Pittsburgh}
  \country{United States}}
\email{jingjingli.cmu@gmail.com}

\author{Yongri Piao}
\correspondingauthor
\affiliation{%
  \institution{Dalian University of Technology}
  \city{Dalian}
  \country{China}}
\email{yrpiao@dlut.edu.cn}

\author{Huchuan Lu}
\affiliation{%
  \institution{Dalian University of Technology}
  \city{Dalian}
  \country{China}}
\email{lhchuan@dlut.edu.cn}

\renewcommand{\shortauthors}{Peng et al.} 

\begin{abstract}
Audio-Visual Instance Segmentation (AVIS) aims to simultaneously classify, segment, and track sounding objects within video sequences. Unlike Audio-Visual Semantic Segmentation (AVS), AVIS involves instance-level modeling across longer video sequences, introducing two key challenges: (1) complex modality-state changes disrupt long-range modeling, and (2) substantial structural and distributional discrepancies between modalities hinder precise instance-level association. Existing methods rely on fixed-step Transformers and recursive Mamba models, lacking adaptability to modality-state changes. In addition, methods performing implicit matching ignore the inherent distributional inconsistencies. To address these issues, we propose a framework with Adaptive Dynamic Step Modulation (ADSM) and Optimal Transport-based Matching Modulation (OT-MM). ADSM adaptively modulates Mamba step sizes using temporal variation, cross-modal discrepancy, and historical context, balancing rapid response to modality-state changes with stable long-range modeling. OT-MM explicitly formulates instance-level cross-modal matching as an entropy-regularized optimal transport problem solved via log-domain Sinkhorn iterations, and further enforces distribution-level coherence with an MMD regularizer. Extensive experiments demonstrate state-of-the-art performance on the AVIS benchmark (+3.76 FSLA, +2.75 HOTA, +2.58 mAP), verified through comprehensive qualitative visualizations. \textit{The code and model are available at \url{https://github.com/happylife-pk/SAMOT}.}
\end{abstract} 



\maketitle
\vspace{-0.4cm}
\begin{figure}[t]
  \centering
  \includegraphics[width=\linewidth]{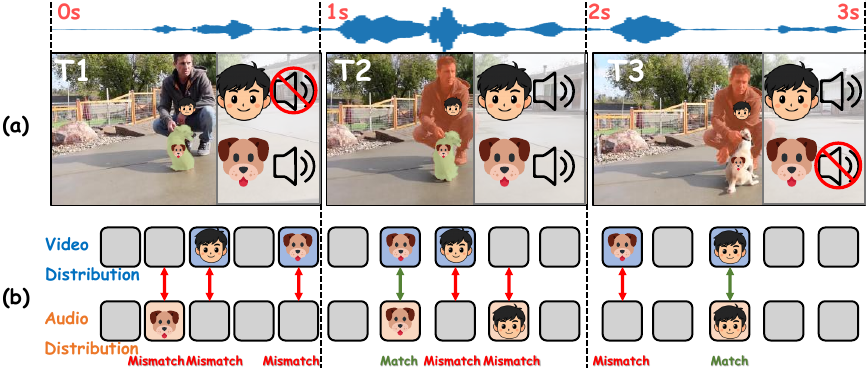}
  \vspace{-0.7cm}
  \caption{Illustration of the two core challenges in AVIS. (a) Although the visual content remains stable, the active sources are the dog at T1, both the person and the dog at T2, and the person
at T3. These modality-state changes disrupt stable long-range modeling. (b) Structural and distributional discrepancies between audio and visual modalities hinder precise instance-level correspondence.}
  \vspace{-0.7cm}
  \label{moti}
\end{figure}
\section{Introduction}
Audio-Visual Instance Segmentation (AVIS)~\cite{guo2025audio} is a novel task designed to simultaneously classify, segment, and track sounding objects at instance level within video sequences by jointly leveraging auditory and visual modalities. This task facilitates a wide range of practical applications, including intelligent surveillance~\cite{xing2025echotraffic}, multimedia content analysis~\cite{deldjoo2018audio}, and video understanding~\cite{cheng2024mixtures}.

Compared to Audio-Visual Semantic Segmentation (AVS), AVIS exhibits two key features: ($i$) it involves longer video sequences and ($ii$) it requires instance-level modeling. These properties introduce two inherent challenges, as shown in Fig.~\ref{moti}. First, long sequences naturally exhibit richer and more complex modality-state changes ($e.g.$, stable visual presence with intermittent audio, or continuous audio presence with intermittent visual absence). Such complex modality-state changes disrupt the stability of long-range modeling. Second, AVIS inherently involves significant mismatches between audio and visual modalities. These mismatches typically manifest as structural and distributional discrepancies, such as the spatial richness of video queries contrasting with the temporal-spectral characteristics of audio queries. Such disparities give rise to pronounced semantic gaps between the two modalities, which complicate the establishment of precise instance-level correspondences.

Existing audio-visual methods~\cite{gao2024avsegformer,yang2024cooperation} typically utilize Transformer-based architectures characterized by multi-head attention mechanisms, where reasoning is uniformly conducted through global computations at fixed time steps. Such fixed temporal processing limits their adaptability to complex modality-state changes in long video sequences. Mamba-based methods~\cite{gong2025avs} attempt to improve long-range dependency modeling through recursive state updates. However, their step-size updates are typically fixed or insufficiently sensitive to input-dependent modality-state alterations, resulting in delayed responses and error propagation in long video sequences.

Concerning modality mismatches, previous methods perform implicit~\cite{li2024multimodal} matching through cross-attention mechanisms~\cite{chen2024cpm} or contrastive learning~\cite{gong2025complementary} to achieve certain progress in alleviating the initial semantic gap for modality fusion. However, these methods often struggle with the inherent inconsistencies between the two modalities, as they primarily focus on learning shared latent representations without explicitly modeling the structural and temporal correlations. This oversight compromises the establishment of multi-modal coherence, causing imprecise matching and fusion.

Based on the above observations, two key questions arise for AVIS: $(1)$ How can a model maintain stable long-range modeling while remaining sensitive to modality-state changes? $(2)$ How can a model explicitly align heterogeneous modalities for precise instance-level associations in complex environments?

To answer these questions, we propose a framework equipped with two core strategies: Adaptive Dynamic Step Modulation (ADSM) and Optimal Transport-based Matching Modulation (OT-MM). First, we introduce ADSM, which dynamically modulates the Mamba step-size during long-sequence modeling. Specifically, ADSM injects audio cues into visual features, estimates temporal variation and cross-modal discrepancy to determine whether the current segment remains stable or is undergoing a modality-state transition. Then, we employ a Delta Gating Router (DGR) to adaptively fuse current visual, audio, and historical cues into a step-size offset, thereby determining the relative contribution of current cross-modal cues and historical context in step-size modulation. In this way, ADSM enables the model to respond more promptly to abrupt modality-state changes while preserving stable long-range propagation.

Second, we introduce OT-MM, a mechanism that explicitly bridges audio-visual queries at the instance level. We formulate instance-level cross-modal matching as an optimal transport problem to establish audio-visual correspondence under structural and temporal mismatches, unlike conventional optimal transport, which relies solely on pairwise matching costs. OT-MM incorporates a Relative Positional Prior (RPP) and a Temporal Consistency Prior (TCP) to capture structural and temporal correspondence, and further employs an MMD regularizer to reduce residual distribution discrepancy. Thus, OT-MM enforces instance-level cross-modal consistency and resolves cross-modal distribution mismatches, facilitating precise associations in multi-instance scenarios. 

Extensive experiments demonstrate that our method achieves state-of-the-art performance on the AVIS benchmark, with improvements of +3.76 FSLA, +2.75 HOTA, and +2.58 mAP. Comprehensive quantitative and qualitative analyses further validate the overall framework and the effectiveness of our key designs for handling complex modality-state changes and cross-modal discrepancies in AVIS scenarios. 

\section{Related Work}
\subsection{Audio-Visual Segmentation}
Related advances in segmentation~\cite{DBLP:journals/tmi/ShiSYWLLGL23,wang2025sam,wang2025personalizing}, grounding and reasoning~\cite{xiao2025towards,liang2026segresearch,gong2025devil,gong2025reinforcing,zhuge2026context}, and long-video and audio-visual understanding~\cite{liang2026long,jiang2026refer,sitong2026vinci2,lu2025learning,lu2026av} provide broader methodological context for Audio-Visual Segmentation (AVS). Notably, AVS has progressed significantly in recent years~\cite{zhou2022audio,li2023catr,hao2024improving,zhou2025audio}. Recent AVS methods~\cite{gao2024avsegformer,yang2024cooperation,ma2024stepping,sun2024unveiling,peng2026selective,gong2025complementary} improve fine-grained cross-modal fusion through audio-conditioned queries, bilateral interaction, and task-specific training strategies. However, AVS typically focuses on semantic segmentation in short videos, without distinguishing multiple instances or maintaining temporal consistency across frames. AVIS~\cite{guo2025audio} extends AVS by incorporating instance-level differentiation and frame-to-frame tracking in long videos. AVISM~\cite{guo2025audio} establishes the AVIS benchmark and employs Transformer-based tracking to preserve sounding-object identities in dynamic multi-instance scenes. Nevertheless, existing methods still struggle with heterogeneous modality dynamics in long sequences.

\subsection{Mamba and Multimodal Modeling}

The Selective State Space Model (Mamba)~\cite{gu2023mamba}, built on structured state space models (SSMs)~\cite{gu2021combining,gu2022parameterization,liu2025defmamba}, provides selective mechanisms for input-dependent state propagation. These mechanisms efficiently capture long-range dependencies while controlling information retention, making Mamba-based methods~\cite{li2024videomamba,erol2024audio} suitable for extended sequences. In audio-visual tasks, AVS-Mamba~\cite{gong2025avs} and AV-Mamba~\cite{huang2024av} integrate temporal and cross-modal selective mechanisms for segmentation and question answering, respectively, supporting efficient spatio-temporal fusion. Mamba has also been applied to audio-visual enhancement~\cite{chao2025leveraging} and continual audio-visual classification~\cite{lin2025mamba}, where sequential modeling improves robustness and alleviates forgetting. Related studies extend temporal and multimodal modeling to speech representation, adaptive fusion, and efficient inference~\cite{DBLP:journals/speech/YeWWXLWCL22,DBLP:conf/ijcai/WenYLXWW022,zhong2026adasurvmamba,liu2025vidcom2,liu2026globalcom2,liu2026mixing,lin2026v-cast}. However, existing methods do not jointly address complex modality-state alterations in long sequences and cross-modal distribution mismatches.

\begin{figure*}[t]
  \centering
  \includegraphics[width=\linewidth,height=0.35\linewidth]{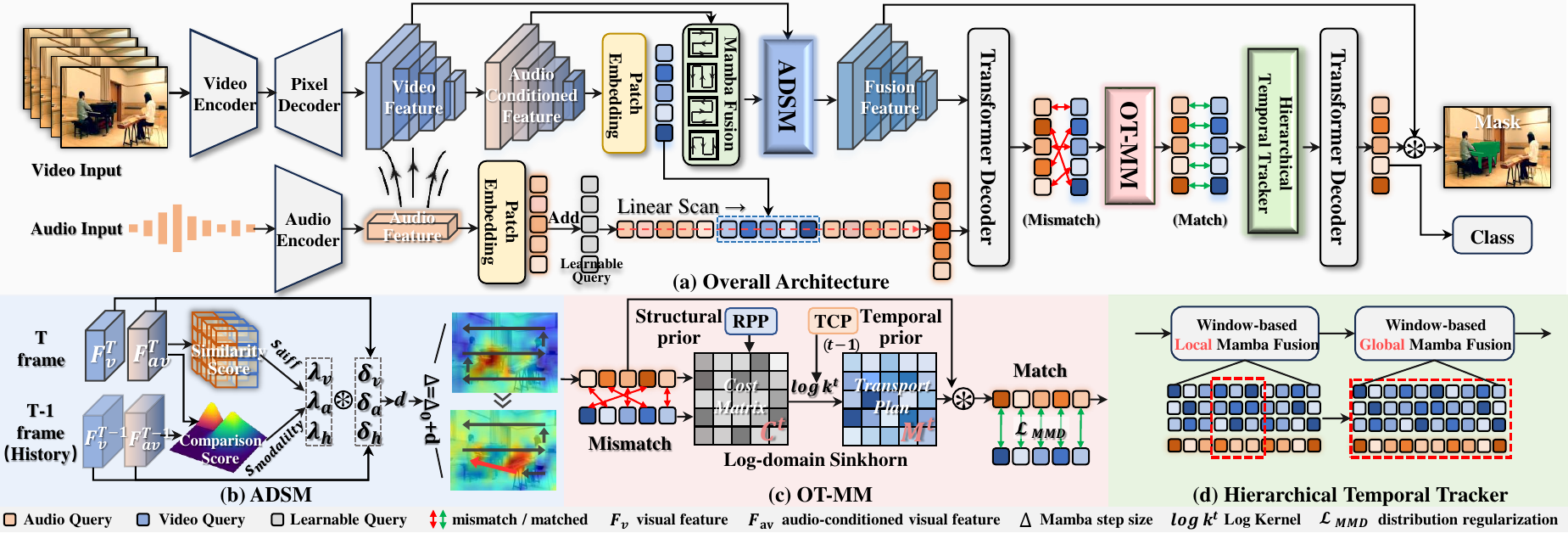}
  \vspace{-0.6cm}
  \caption{(a) Overall architecture of the proposed framework. (b) ADSM dynamically modulates Mamba step sizes using temporal variation, cross-modal discrepancy, and the Delta Gating Router (DGR). (c) OT-MM performs instance-level audio-visual matching with a Relative Positional Prior (RPP) and a Temporal Consistency Prior (TCP), followed by MMD regularization. (d) The Hierarchical Temporal Tracker performs sliding-window refinement and global temporal modeling for final tracking.}
  \vspace{-10pt}
  \label{DAMF}
\end{figure*}

\section{Method}
\subsection{Overview}
As shown in Fig.~\ref{DAMF}(a), given a video sequence $V \in \mathbb{R}^{T\times3\times H \times W}$ with $T$ frames and its corresponding audio clip, we adopt ResNet~\cite{he2016deep} as the visual backbone to extract multi-scale spatial features from video frames. The visual encoder produces hierarchical feature maps, which are further refined by a pixel decoder~\cite{zhudeformable} to generate high-resolution embeddings $F_v=[f_{v}^{i}]^{4}_{i=1}$, where $f_{v}^{i} \in \mathbb{R}^{T\times C\times H_i \times W_i}$ and $H_i \times W_i$ denotes the spatial resolution of the $i$-th stage. Meanwhile, the input audio is converted into a mono mel-spectrogram and encoded by a pre-trained VGGish model~\cite{hershey2017cnn} into an audio feature sequence $f_{a} \in \mathbb{R}^{T \times C_{a}}$, where $C_{a}$ is the audio feature dimension.

Based on $F_v$ and $f_a$, the frame-level localizer proceeds along two complementary branches. On the visual side, we inject audio cues into the lower three visual scales $[f_v^i]_{i=1}^{3}$ and apply Adaptive Dynamic Step Modulation (ADSM) to dynamically regulate the visual state update, yielding audio-conditioned visual features $[f_{av}^{i}]_{i=1}^{3}$ that are more sensitive to modality-state changes. In parallel, the audio features are projected into query embeddings and further associated with the highest-level visual tokens $f_v^4$ to form audio queries $q_a \in \mathbb{R}^{T \times N_a \times C}$, where $N_a$ denotes the number of audio queries. The Transformer decoder then takes $[f_{av}^{i}]_{i=1}^{3}$ and $q_a$ as inputs to derive instance-level visual queries $q_v \in \mathbb{R}^{T \times N_v \times C}$, where $N_v$ is the number of frame queries.

At the video level, Optimal Transport-based Matching Modulation (OT-MM) explicitly aligns $q_a$ and $q_v$ by solving a log-domain Sinkhorn transport problem and further regularizing the aligned representations with an MMD loss, thereby producing matched audio queries $\tilde q_a$ and establishing robust instance-level cross-modal correspondence. Finally, the Hierarchical Temporal Tracker performs sliding-window local refinement followed by full-sequence Mamba modeling over $q_v$ and $\tilde q_a$, thereby preserving short-term consistency while capturing long-range temporal dependencies for final segmentation and tracking predictions.

\subsection{Adaptive Dynamic Step Modulation}
To effectively capture complex modality-state alterations in long-range audio-visual sequences, we propose the Adaptive Dynamic Step Modulation (ADSM) mechanism based on the Mamba architecture. Unlike conventional methods that utilize fixed step-sizes, ADSM dynamically adjusts the step-size according to both current cross-modal states and historical context, striking a balance between immediate response and long-range modeling stability.

We first inject audio cues into multi-scale visual features to obtain audio-conditioned visual features, enabling the visual stream to reflect the current sounding state. Specifically, the audio feature $f_a^t$ is fused with the visual representation $f_v^{i,t}$ at scale $i$ and  time step $t$ via:
\begin{equation}
    f_{av}^{i,t} = f_v^{i,t}\odot\left[1+\tanh\left(\mathrm{CrossAttn}(f_v^{i,t}, f_{a}^t)\right)\cdot\alpha\right],
\end{equation}
where $\odot$ denotes element-wise multiplication, and $\alpha$ is a learnable scaling parameter controlling the intensity of cross-modal fusion. For brevity, we omit the scale index $i$ in the following equations and use $f_v^t$ and $f_{av}^t$ to denote the corresponding current visual and audio-conditioned visual representations.

As shown in Fig.~\ref{DAMF}(b), to determine whether the current segment remains stable or is undergoing a modality-state transition, ADSM estimates two complementary signals: a temporal variation score $s_{\text{modality}}$, which captures the abruptness of state changes across adjacent time steps, and a cross-modal discrepancy score $s_{\text{diff}}$, which measures the semantic deviation of the audio-conditioned visual representation from the raw visual representation. The temporal variation score is defined as:
\begin{equation}
    s_{\text{modality}} = \frac{\max \left( \lVert f_{av}^t - f_{av}^{t-1} \rVert_{\infty}, \lVert f_{v}^t - f_{v}^{t-1} \rVert_{\infty} \right)}{\max \left( \lVert f_{av}^t \rVert_{\infty}, \lVert f_{v}^t \rVert_{\infty} \right)}.
\end{equation}
In this formulation, the Max norm highlights the most pronounced temporal variation, ensuring that ADSM remains highly sensitive to abrupt state transitions. Complementary to this, the cross-modal discrepancy score is defined as:
\begin{equation}
    s_{\text{diff}} = 1 - \frac{f_{av}^t \cdot f_{v}^t}{\lVert f_{av}^t \rVert_2 \lVert f_{v}^t \rVert_2}.
\end{equation}
By utilizing a cosine-based distance, this metric focuses on cross-modal semantic deviation, enabling the detection of substantial cross-modal inconsistencies at the current step. Consequently, $s_{\text{modality}}$ captures temporal abruptness, whereas $s_{\text{diff}}$ reflects the degree of instantaneous cross-modal conflict. Together, they characterize whether the current segment is temporally stable and whether its audio-visual evidence remains cross-modally consistent.

However, perceiving modality-state alterations alone is insufficient; the model must further determine which source of evidence should dominate the step-size adjustment. To this end, we introduce the Delta Gating Router (DGR), which decomposes the update into visual, audio, and history branches and adaptively weights them according to the current state indicators. We then estimate the base modulation influences of the three branches, denoted as $\delta_v$, $\delta_a$, and $\delta_h$, which respectively capture current audio-conditioned visual cues, current audio cues, and historical cross-modal context:
\begin{equation}
\begin{gathered}
\delta_v = \tanh(\mathrm{MLP}(f^t_{av})),
\quad \delta_a = \tanh(\mathrm{MLP}(f^t_a)), \\
\delta_h = \tanh(\mathrm{MLP}(f^{t-1}_{av} + f^{t-1}_{a})).
\end{gathered}
\end{equation}
where MLP denotes a multi-layer perceptron. Guided by the perceived state indicators, the router assigns adaptive weights $\lambda_v$, $\lambda_a$, and $\lambda_h$ to these branches:
\begin{equation}
    [\lambda_v, \lambda_a, \lambda_h] = \text{softmax} \left( \text{MLP} \left( [s_\text{diff}, s_\text{modality}] \right) \right).
\end{equation}
This adaptive routing strategy ensures that when state changes are abrupt or cross-modal discrepancy is high, the model emphasizes immediate modality-specific evidence, whereas under smoother dynamics, it places more weight on historical context to maintain long-range continuity.

Finally, this routing decision is translated into a concrete step-size offset. The weighted fusion of the three branches yields an adaptive offset $d$, which dynamically modulates the original fixed step-size $\Delta_0$ to obtain the adjusted step-size $\Delta$:
\begin{equation}
\begin{gathered}
    d = \lambda_v\delta_v + \lambda_a\delta_a + \lambda_h\delta_h, \\
    \Delta = \Delta_0 + d.
\end{gathered}
\end{equation}
Crucially, the adjusted $\Delta$ is directly injected into the discretization step of the selective state-space mechanism, thereby modulating the update rate of the Mamba dynamics. Thus, ADSM enables rapid adaptation under abrupt modality-state shifts while preserving stable long-range propagation in temporally smoother segments.
\begin{table*}[t]
  \centering
  \captionsetup{labelfont=bf, font=small} 
  \caption{Performance comparison of different models on VIS, AVS, and AVIS tasks. The \colorbox{best}{red} and \colorbox{second}{blue} backgrounds denote the best and second-best results on AVIS tasks. \textsuperscript{*} indicates models trained with an MS-COCO pre-trained visual backbone.}
  \vspace{-6pt}
  \label{tab:performance_comparison}
  
  \renewcommand{\arraystretch}{0.85}
  
  \begin{tabular*}{0.96\textwidth}{@{\hspace{1em}\extracolsep{\fill}} l l c c c c c c c c @{\hspace{1em}}}
    \toprule
    \textbf{Task} & \textbf{Model} & \textbf{Reference} & \textbf{Audio} & \textbf{FSLA} & \textbf{HOTA} & \textbf{mAP} & \textbf{FSLAn} & \textbf{FSLAs} & \textbf{FSLAm} \\
    \midrule
    
    \multirow{5}{*}{VIS}
    & TeViT~\cite{yang2022temporally}          & CVPR'22   & \ding{55} & 32.28 & 53.67 & 31.52 & 0.00 & 28.07 & 39.18 \\
    & SeqFormer~\cite{wu2022seqformer}      & ECCV'22   & \ding{55} & 30.32 & 54.32 & 32.79 & 25.03 & 21.76 & 36.46 \\
    & VITA~\cite{heo2022vita}           & NeurIPS'22   & \ding{55} & 38.04 & 57.48 & 36.25 & 15.04 & 27.98 & 47.45 \\
    & DAVIS~\cite{zhang2023dvis}          & ICCV'23   & \ding{55} & 23.99 & 49.12 & 19.83 & 14.61 & 24.83 & 24.69 \\
    & LBVQ~\cite{fang2024learning}           & TCSVT'24  & \ding{55} & 34.73 & 56.97 & 36.58 & 27.71 & 29.52 & 38.96 \\
    
    \midrule
    
    \multirow{4}{*}{AVS}
    & AVSegFormer~\cite{gao2024avsegformer}    & AAAI'24   & \ding{51} & 35.66 & 55.74 & 35.72 & 18.58 & 27.51 & 43.08 \\
    & COMBO~\cite{yang2024cooperation}          & CVPR'24   & \ding{51} & 39.49 & 57.39 & 37.84 & 21.91 & 27.18 & 49.63 \\
    & VCT~\cite{huang2025revisiting}             & CVPR'25   & \ding{51} & 41.89 & 60.71 & 40.06 & 32.17 & 27.44 & 52.01 \\
    & SDAVS~\cite{peng2026selective}          & TMM'26    & \ding{51} & 41.50 & 60.30 & 39.80 & 31.08 & 28.31 & 51.46 \\
    
    \midrule
    
    \multirow{3}{*}{AVIS}
    & AVISM~\cite{guo2025audio}          & CVPR'25   & \ding{51} & 42.78 & 61.73 & 40.57 & \cellcolor{second}32.22 & 29.83 & \cellcolor{second}52.40 \\
    & ACVIS~\cite{seo2026learning}          & ICASSP'26 & \ding{51} & \cellcolor{second}42.87 & \cellcolor{second}62.09 & \cellcolor{second}42.14 & 21.16 & \cellcolor{second}30.62 & 52.34 \\
    & Ours           & ---       & \ding{51} & \cellcolor{best}46.07 & \cellcolor{best}62.58 & \cellcolor{best}42.85 & \cellcolor{best}37.81 & \cellcolor{best}32.46 & \cellcolor{best}55.81 \\
    
    \midrule
    
    \multirow{3}{*}{AVIS\textsuperscript{*}}
    & AVISM\textsuperscript{*}~\cite{guo2025audio} & CVPR'25   & \ding{51} & 44.42 & 64.52 & 45.04 & \cellcolor{second}20.62 & 32.62 & 54.99 \\
    & ACVIS\textsuperscript{*}~\cite{seo2026learning}                     & ICASSP'26 & \ding{51} & \cellcolor{second}46.48 & \cellcolor{second}65.12 & \cellcolor{second}46.68 & 10.74 & \cellcolor{best}34.45 & \cellcolor{second}58.81 \\
    & Ours\textsuperscript{*}  & ---       & \ding{51} & \cellcolor{best}48.18 & \cellcolor{best}67.27 & \cellcolor{best}47.62 & \cellcolor{best}36.73 & \cellcolor{second}32.87 & \cellcolor{best}60.68 \\
    
    \bottomrule
  \end{tabular*}
  
  \vspace{-10pt}
\end{table*}
\subsection{Optimal Transport-based Matching Modulation}
As shown in Fig.~\ref{DAMF}(c), we introduce the Optimal Transport-based Matching Modulation (OT-MM) mechanism to address the inherent semantic mismatches between heterogeneous modalities and to construct robust instance-level correspondences. Unlike standard implicit cross-attention, OT-MM explicitly formulates the cross-modal matching as an optimal transport problem uniquely constrained by spatial-temporal priors, thereby establishing robust instance-level cross-modal correspondence for subsequent temporal tracking.

Let $q^t_a \in \mathbb{R}^{N_a\times C}$ and $q^t_v \in \mathbb{R}^{N_v\times C}$ denote the temporally-sampled audio and spatially-sampled video embeddings at time step $t$. The goal is to find a soft transport plan $M^t \in \mathbb{R}^{N_a\times N_v}$ that minimizes the assignment cost. To explicitly restrict the search space of the transport plan and encourage the model to prioritize structurally corresponding queries, we incorporate a Relative Positional Prior (RPP) into the transport cost. This effectively prevents distinctive audio signals from being spuriously matched to noisy or homogeneous visual backgrounds. Assuming $\ell_2$-normalized features, the cost matrix $C^t$ is defined as:
\begin{equation}
    C^t_{i,j} = \left(1 - \langle q_a^t[i], q_v^t[j] \rangle \right) + \beta \frac{|i-j|}{\max(N_a, N_v)},
\end{equation}
where $\beta$ controls the intensity of the relative positional prior. This relative positional prior encourages the matching of queries that are proximate in the unrolled feature sequences, effectively reducing false-positive global associations. Under this cost mechanism, the cross-modal matching is naturally formulated as an entropy-regularized optimal transport problem, in which the transport plan is optimized under uniform marginal constraints to produce a balanced correspondence matrix. 

While $C^t$ effectively resolves intra-frame structural ambiguities, the AVIS task additionally requires maintaining the temporal consistency of sounding objects across extended video sequences. To achieve this inter-frame stability without introducing additional learnable parameters, we integrate a Temporal Consistency Prior (TCP) by recycling the transport plan from the previous time step. We define the temporally-smoothed log-kernel matrix $\log K^t$ as:
\begin{equation}
\log K^t =
\begin{cases}
-\frac{C^t}{\epsilon} + \gamma \log\left(M^{t-1} + \eta\right), & t > 1,\\
-\frac{C^t}{\epsilon}, & t = 1,
\end{cases}
\end{equation}
where $\epsilon$ is the entropic regularization coefficient, $\gamma$ acts as a momentum factor dictating the influence of the historical transport prior, and $\eta$ is a small constant to prevent invalid logarithmic operations. 

During standard Sinkhorn iterations, numerical underflow frequently destabilizes gradient propagation. To ensure robust numerical stability and continuous gradient flow, we compute the marginal projections strictly in the logarithmic domain using the Log-Sum-Exp ($\operatorname{LSE}$) operator. \emph{Detailed derivations of the alternating updates are provided in the supplementary material.} After $L$ Sinkhorn iterations, the converged dual scaling variables (denoted as $f_i^{(L)}$ and $g_j^{(L)}$ for the audio and visual modalities, respectively) are used to map the log-kernel back to the probability simplex, yielding the final transport plan $M^t \in \mathbb{R}^{N_a \times N_v}$:
\begin{equation}
    M^t_{i,j} = \exp \left( f_i^{(L)} + g_j^{(L)} + \log K^t(i,j) \right).
\end{equation}

This approach guarantees robust numerical stability and continuous gradient flow throughout the entire optimization process, culminating in a highly reliable multimodal matching. The final matched audio representation $\tilde q^t_a$ is derived via a differentiable transport operation:
\begin{equation} 
\tilde q^t_a = {{(M^t)}^{\mathsf{T}}}q^t_a,
\end{equation} 
where $\mathsf{T}$ denotes the transpose operation, projecting the audio representation into the video embedding domain. To enhance the robustness of this match, the OT module operates iteratively across consecutive temporal steps, forming local-to-global consistency in feature space, ensuring that the established correspondence is both temporally smooth and semantically precise.

Furthermore, to effectively constrain modality gaps remaining after OT matching and further enforce distribution-level coherence, we employ a Maximum Mean Discrepancy (MMD)-based distributional loss, regularizing the overall audio-video feature distributions in a Reproducing Kernel Hilbert Space (RKHS). The MMD loss is defined as:
\begin{equation}
\begin{gathered}
\mathcal{L}_{\mathrm{MMD}} =
\frac{1}{N_a^2}\sum_{i=1}^{N_a}\sum_{i'=1}^{N_a} k\big(\tilde q_a^t[i], \tilde q_a^t[i']\big)
+\\
\frac{1}{N_v^2}\sum_{j=1}^{N_v}\sum_{j'=1}^{N_v} k\big(q_v^t[j], q_v^t[j']\big)
-
\frac{2}{N_a N_v}\sum_{i=1}^{N_a}\sum_{j=1}^{N_v} k\big(\tilde q_a^t[i], q_v^t[j]\big),\\
k(x,y)=\frac{1}{|S|}\sum_{\sigma\in S}\exp\left(-\frac{\|x-y\|_2^2}{2\sigma^2}\right),
\end{gathered}
\end{equation}
where $k(x,y)$ is a multi-kernel Gaussian function, in which each bandwidth $\sigma$ is selected from a fixed multi-scale set $S=\{0.1,1.0,10.0\}$. We adopt this design to capture cross-modal discrepancy at different similarity ranges, \emph{while additional analysis on fixed-bandwidth ablations is provided in the supplementary material.}

\subsection{Hierarchical Temporal Tracker}
To track sounding objects across video sequences, we employ a hierarchical temporal tracker that combines local temporal refinement with full-sequence Mamba modeling. As shown in Fig.~\ref{DAMF}(d), we first apply a sliding-window refinement module with window size $w=3$ and stride 1 to preserve short-term consistency, where each local segment spans frames $t:t+w-1$. For each window, we extract the corresponding visual object queries $q_v^{t:t+w-1}$ and matched audio queries $\tilde q_a^{t:t+w-1}$. Bidirectional scanning is then performed within the window, followed by a lightweight fusion layer for cross-modal interaction and local temporal refinement:
\begin{equation}
\begin{gathered}
q^{t:t+w-1}_{av} = M_v(q^{t:t+w-1}_v) \odot M_a(\tilde q^{t:t+w-1}_a), \\
\hat q^{t:t+w-1}_{av} = M_v(q^{t:t+w-1}_v) + M_a(\tilde q^{t:t+w-1}_a), \\
\tilde q^{t:t+w-1}_v = \sigma(GAP(M_{av}(q^{t:t+w-1}_{av}))) \odot \hat q^{t:t+w-1}_{av} + q^{t:t+w-1}_v,
\end{gathered}
\end{equation}
where $M_{v/a/av}$ denote Mamba blocks, $\sigma$ denotes the Sigmoid function, and $GAP$ denotes global average pooling along the temporal dimension. This local refinement improves short-term tracking consistency and alleviates drift in appearance or sound correlation.

After window-based refinement, we further apply Mamba over the entire sequence to model global temporal dependencies and produce the final temporal query embeddings, thereby complementing the preceding local refinement. In this way, the tracker preserves local consistency while incorporating long-range contextual cues for object motion, sound-source continuity, and extended temporal dependency modeling.
\begin{figure*}[ht]
  \centering
  \includegraphics[width=0.9\linewidth]{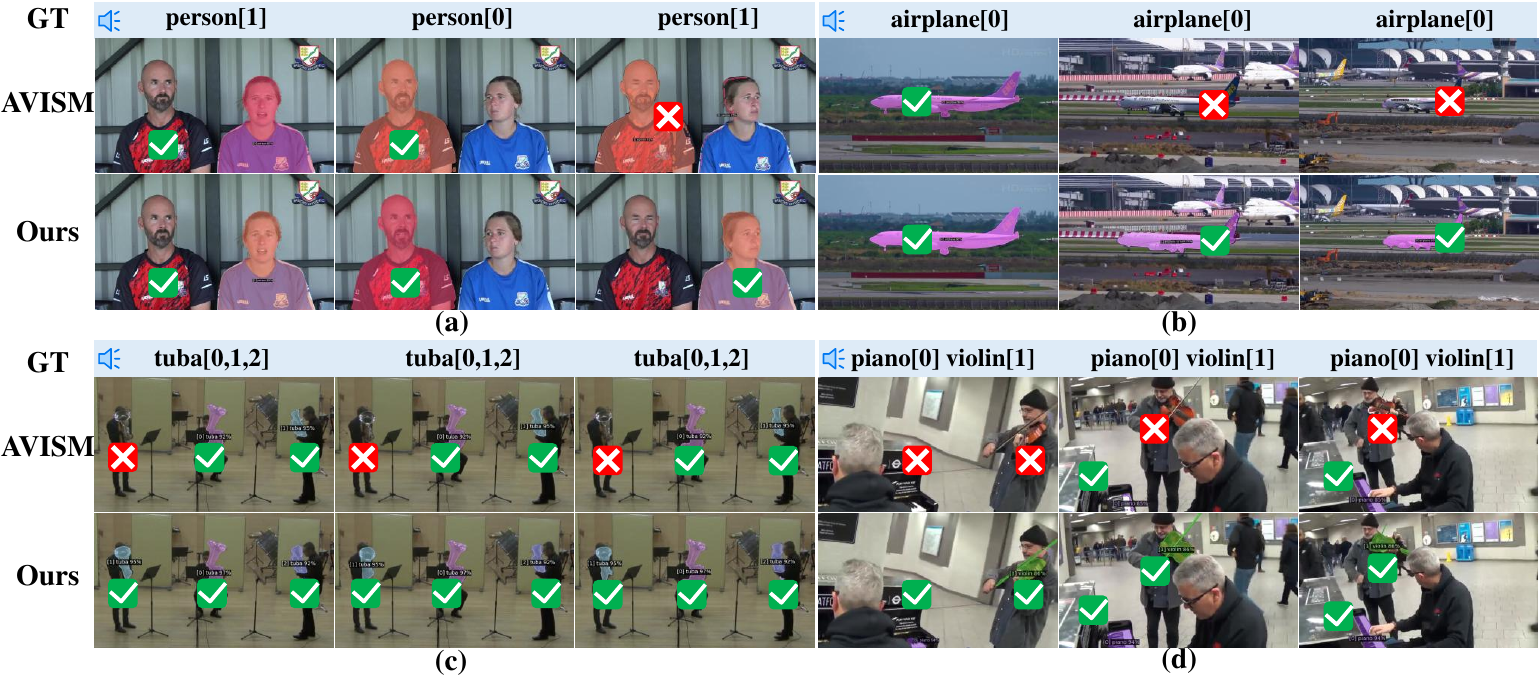}
  \vspace{-0.3cm}
  \caption{Qualitative comparison between AVISM and Ours across diverse audio scenarios with varying sound sources.}
  \label{avis}
  \vspace{-10pt}
\end{figure*}

\section{Experiments}
\subsection{Overall Comparisons}
\noindent\textbf{Dataset and Evaluation Metrics.} We evaluate our method on AVISeg~\cite{guo2025audio}, the first benchmark for audio-visual instance segmentation, whose videos have an average duration of 61.4 seconds. Following~\cite{guo2025audio}, we report mean Average Precision (mAP), Higher Order Tracking Accuracy (HOTA), and Frame-Level Sound Localization Accuracy (FSLA). \emph{Details of the dataset and evaluation metrics are provided in the supplementary material.}

\noindent\textbf{Training Details.} For fair comparison, all methods use the same input resolution and ResNet-50 backbone. Our model employs a ResNet-50~\cite{he2016deep} visual backbone, a VGGish~\cite{hershey2017cnn} audio encoder pretrained on AudioSet~\cite{gemmeke2017audio}, and the query-learning protocol of Mask2Former~\cite{cheng2022masked} and VITA~\cite{heo2022vita}. We train the model for 50,000 iterations with batch size 1 using AdamW and a stepwise learning-rate schedule on NVIDIA RTX 4090 GPUs. We set $\alpha=0.2$, $\beta=0.1$, $\epsilon=0.01$, $\gamma=0.1$, $L=10$, and $\eta=10^{-8}$; the remaining Mamba settings follow VMamba~\cite{liu2024vmamba}. The objective follows AVISM~\cite{guo2025audio} with an additional $\mathcal{L}_{\mathrm{MMD}}$, and all loss weights are 1.0. Further implementation details are provided in the supplementary material.

\noindent\textbf{Comparison Protocol.} We compare our method with representative video instance segmentation (VIS) methods~\cite{cheng2021mask2formervideoinstancesegmentation,yang2022temporally,wu2022seqformer,heo2022vita,zhang2023dvis,fang2024learning}, AVS methods~\cite{gao2024avsegformer,yang2024cooperation,shen2025frequency,huang2025revisiting,peng2026selective}, and the AVIS baseline AVISM~\cite{guo2025audio}. Following AVISM~\cite{guo2025audio}, VIS and AVS methods are adapted to the AVIS benchmark under the same backbone, input resolution, and training protocol. The results are summarized in Table~\ref{tab:performance_comparison}. It is worth noting that the AVISM result reported in Table~\ref{tab:performance_comparison} corresponds to the official benchmark baseline, whereas the baseline used in our later ablation studies is our reproduced framework equipped with the hierarchical temporal tracker but without ADSM or OT-MM. Therefore, the two baselines are not numerically identical.

As shown in Table~\ref{tab:performance_comparison}, our method achieves the best primary metrics under both standard and MS-COCO-pretrained settings. Under the standard setting, it surpasses AVISM by 3.29 FSLA, 0.85 HOTA, and 2.28 mAP; with MS-COCO pretraining, the corresponding gains are 3.76, 2.75, and 2.58, respectively. The FSLA improvement indicates more accurate sounding-object localization in long and dynamic sequences, while the HOTA and mAP gains reflect stronger temporal association and instance-level segmentation. Notably, our method performs better in challenging silent and multi-source scenarios, as evidenced by the improvements in FSLAn and FSLAm, verifying the effectiveness of our design in handling complex modality-state changes and cross-modal discrepancies in AVIS.

\subsection{Qualitative and Mechanism Analysis}
\noindent\textbf{How does the framework perform in challenging scenarios?} 

To intuitively demonstrate the effectiveness of our framework, we present qualitative comparisons between Ours and AVISM in challenging AVIS scenarios, as shown in Fig.~\ref{avis}. The displayed frames are adjacent samples selected to highlight inter-frame modality-state changes and complex multi-instance interactions, and GT denotes the currently sounding object instances.

In Fig.~\ref{avis}(a-b), the dominant challenge lies in complex modality-state changes. In Fig.~\ref{avis}(a), the visual content remains nearly unchanged while the active sound source alternates across different persons. In Fig.~\ref{avis}(b), the sounding aircraft remains consistent while the visible target undergoes large motion and background variation. In both cases, AVISM fails to promptly perceive changes in the current modality state and therefore cannot adapt its temporal modeling accordingly. In contrast, our method can more accurately track and segment the currently sounding object, demonstrating the effectiveness of ADSM in stabilizing temporal modeling while remaining responsive to modality-state transitions.

In Fig.~\ref{avis}(c-d), we further consider multi-instance scenes with substantial cross-modal mismatch. When multiple instruments sound simultaneously, AVISM produces imprecise correspondence and localization, whereas our method establishes clearer instance-level associations and more accurate masks. These results demonstrate that OT-MM improves cross-modal matching in acoustically and visually complex scenarios.
\begin{figure}[t]
  \centering
  \includegraphics[width=1\linewidth]{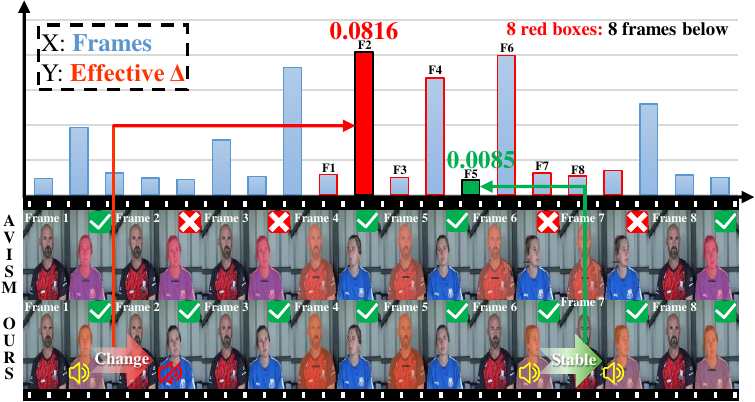}
  \vspace{-0.66cm}
  \caption{Visualization of the effective step size $\Delta$ in ADSM. The upper bar chart shows frame-wise $\Delta$ values, and the lower rows present the corresponding predictions of AVISM and Ours on adjacent frames.}
  \vspace{-15pt}
  \label{ADSM}
\end{figure}

\noindent\textbf{How does ADSM respond to modality-state changes?} 

To provide a more intuitive understanding of ADSM, we visualize the effective step size $\Delta$ together with representative frames exhibiting different modality-change states, as shown in Fig.~\ref{ADSM}. During temporally stable segments, the effective $\Delta$ remains at a relatively low level, indicating that the model preserves historical information and performs smoother state propagation. When the sounding object changes or the modality state undergoes abrupt transition, the effective $\Delta$ rises noticeably, allowing the model to update its hidden state more aggressively and quickly adapt to the new audio-visual configuration. This behavior is consistent with the design motivation of our ADSM, retaining long-range stability in steady segments while improving responsiveness under sudden modality-state alterations.

\begin{figure}[t]
  \centering
  \includegraphics[width=\linewidth]{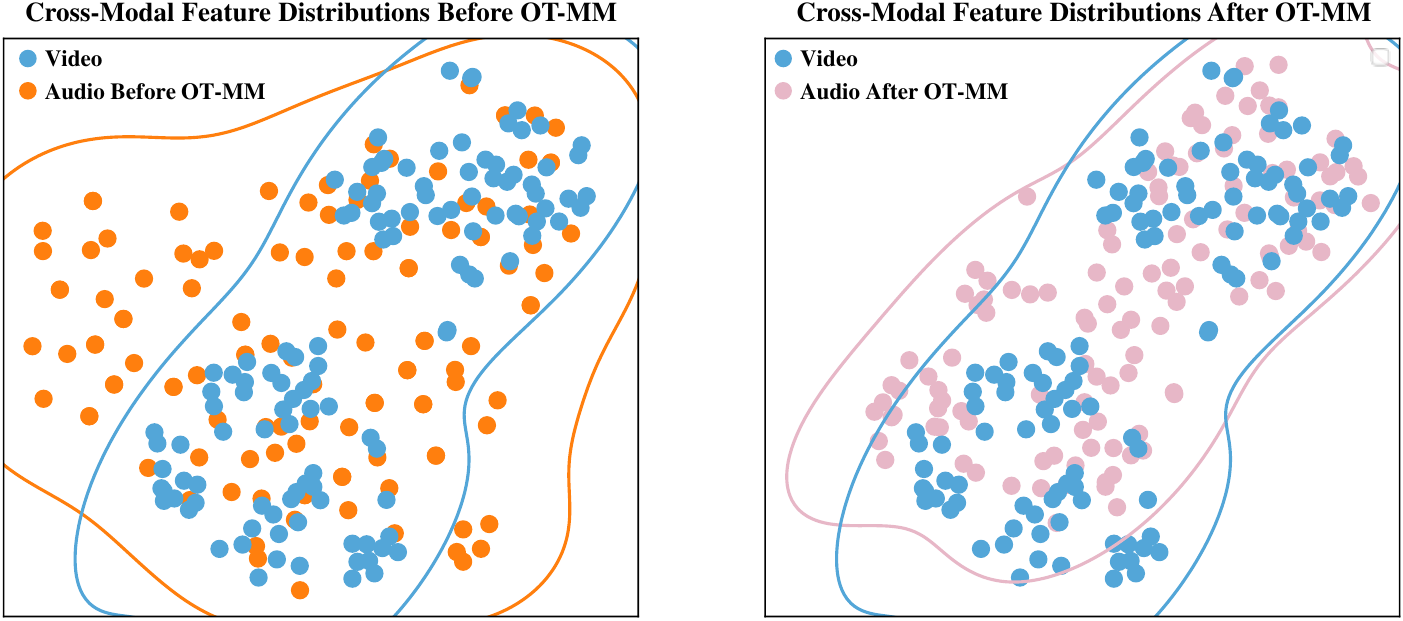}
  \vspace{-0.6cm}
  \caption{Visualization of cross-modal feature distributions before and after OT-MM. (a) Before OT-MM, the audio (orange) and video (blue) embeddings exhibit large distributional divergence. (b) After OT-MM, the matched audio (pink) shows strong overlap with the video distribution.}
\vspace{-10pt}
  \label{OT-MMM}
\end{figure}

\begin{figure}[t]
  \centering
  \includegraphics[width=\linewidth]{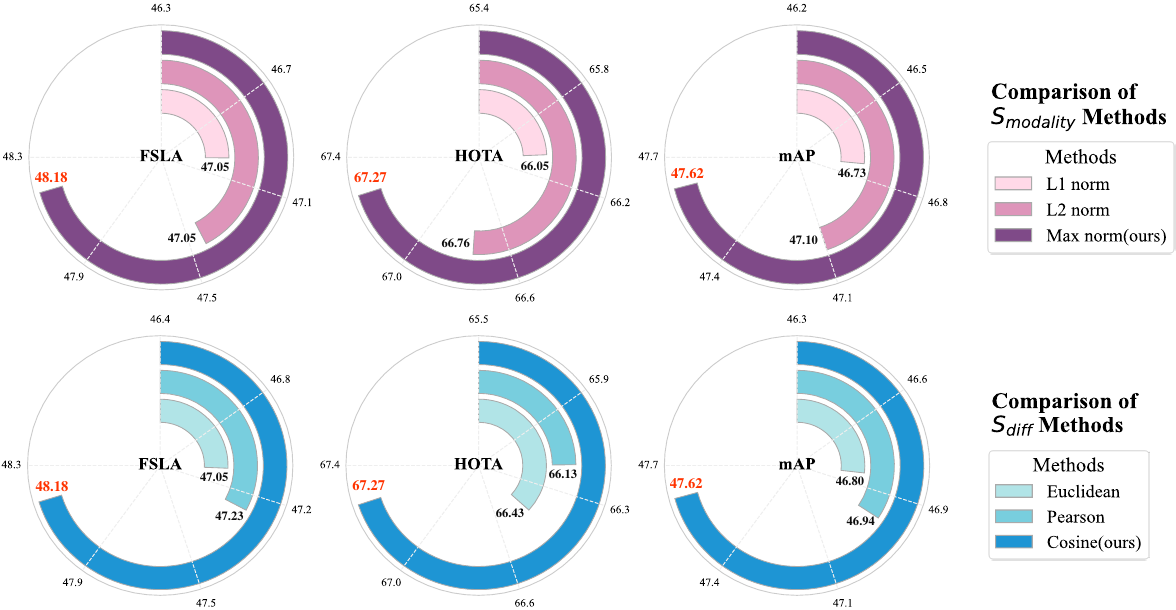}
  \vspace{-0.6cm}
  \caption{Comparison of alternative distance formulations for $s_{\text{modality}}$ and $s_{\text{diff}}$ in ADSM. The selected max-norm and cosine formulations achieve the best performance.}
  \vspace{-15pt}
  \label{cycle}
\end{figure}
\noindent\textbf{How does OT-MM improve cross-modal matching?}

To further reveal how OT-MM improves cross-modal matching, we visualize the audio-video feature distributions before and after OT-MM using t-SNE~\cite{maaten2008visualizing} with Kernel Density Estimation (KDE)~\cite{zheng2013quality}, as shown in Fig.~\ref{OT-MMM}. Before matching, the audio and visual embeddings form largely separated clusters, reflecting substantial structural and distributional discrepancies between the two modalities. After OT-MM, the matched audio embeddings become significantly closer to the visual embeddings, and the corresponding density contours exhibit much stronger overlap. These results indicate that OT-MM does not merely fuse heterogeneous features, but instead establishes soft instance-level correspondence through transport-based matching while reducing residual distribution gaps, thereby producing more coherent multimodal representations for downstream segmentation and tracking.

\subsection{Ablation Studies}
\noindent\textbf{Overall module ablation.}
To evaluate the contribution of the two core strategies in our framework, we progressively introduce ADSM and OT-MM into the reproduced baseline. As shown in Table~\ref{tab:ablation_study}, adding either ADSM or OT-MM consistently improves FSLA, HOTA, and mAP over the baseline, verifying that both temporal step modulation and explicit cross-modal matching are beneficial for AVIS. In particular, ADSM yields larger gains in FSLA, indicating its effectiveness in handling complex modality-state changes during long-range sequence modeling, while OT-MM provides more stable improvements in HOTA and mAP by enhancing instance-level cross-modal correspondence. When the two modules are jointly applied, the model achieves the best overall performance, demonstrating that ADSM and OT-MM play complementary roles in improving temporal modeling and multimodal alignment.
\begin{table}[t]
    \centering
    \captionsetup[subtable]{skip=2pt}
    \caption{Ablation study of ADSM and OT-MM modules.}
    \vspace{-7pt}
    \label{tab:ablation_study}
    \resizebox{0.72\linewidth}{!}{
    \begin{tabular}{cccccc}
    \hline
    Index & ADSM & OT-MM & FSLA & HOTA & mAP \\
    \hline
    (1) &  &  & 45.56 & 65.53 & 46.49 \\
    (2) & \ding{51} &  & 47.29 & 65.99 & 47.05 \\
    (3) &  & \ding{51} & 46.64 & 66.53 & 46.99 \\
    \rowcolor{best}
    (4) & \ding{51} & \ding{51} & \textbf{48.18} & \textbf{67.27} & \textbf{47.62} \\
    \hline
    \end{tabular}}
    \vspace{-10pt}
\end{table}
\begin{table}[t]
    \centering
    \captionsetup[subtable]{skip=2pt}
    \caption{Ablation study on ADSM. SP denotes state perception branch implemented by the joint use of $s_{\text{modality}}$ and $s_{\text{diff}}$.}
    \vspace{-7pt}
    \label{tab:ablation_study_adsm}
    \resizebox{0.72\linewidth}{!}{
    \begin{tabular}{cccccc}
    \hline
    Index & SP & DGR & FSLA & HOTA & mAP \\
    \hline
    (1) &  &  & 46.64 & 66.53 & 46.99 \\
    (2) & \ding{51} &  & 47.08 & 66.78 & 47.03 \\
    (3) &  & \ding{51} & 47.21 & 66.92 & 47.17 \\
    \rowcolor{best}
    (4) & \ding{51} & \ding{51} & \textbf{48.18} & \textbf{67.27} & \textbf{47.62} \\
    \hline
    \end{tabular}}
    \vspace{-10pt}
\end{table}
\begin{table}[t]
    \centering
    \captionsetup[subtable]{skip=2pt}
    \caption{Ablation study on OT-MM, where RPP and TCP denote the relative positional prior and temporal consistency prior, respectively.}
    \vspace{-7pt}
    \label{tab:ablation_study_otmm}
    \resizebox{0.72\linewidth}{!}{
    \begin{tabular}{cccccc}
    \hline
    Index & RPP & TCP & FSLA & HOTA & mAP \\
    \hline
    (1) &  &  & 47.29 & 65.99 & 47.05 \\
    (2) & \ding{51} &  & 47.71 & 66.31 & 47.24 \\
    (3) &  & \ding{51} & 47.54 & 66.87 & 47.19 \\
    \rowcolor{best}
    (4) & \ding{51} & \ding{51} & \textbf{48.18} & \textbf{67.27} & \textbf{47.62} \\
    \hline
    \end{tabular}}
    \vspace{-10pt}
\end{table}
\begin{table}[t]
    \centering
    \captionsetup[subtable]{skip=2pt}
    \caption{Comparison of different Mamba scanning mechanisms and OT-based matching strategies.}
    \vspace{-8pt}
    \label{tab:scan_ot_comparison}
    \resizebox{0.98\linewidth}{!}{
    \begin{tabular}{|ccc|ccc|ccc|}
    \hline
    \multicolumn{3}{|c|}{\textbf{Mamba Scan}} & \multicolumn{3}{c|}{\textbf{OT Matching}} & \multicolumn{3}{c|}{\textbf{Metrics}} \\
    \hline
    \textbf{Std. Mamba} & \textbf{SS2D} & \textbf{ADSM} & \textbf{OT-LA} & \textbf{Sinkhorn} & \textbf{OT-MM} & \textbf{FSLA} & \textbf{HOTA} & \textbf{mAP} \\
    \hline
    \ding{51} &  &  &  &  & \ding{51} & 45.64 & 65.09 & 43.98 \\
     & \ding{51} &  &  &  & \ding{51} & 46.36 & \cellcolor{second}66.52 & \cellcolor{second}45.33 \\
     &  & \ding{51} & \ding{51} &  &  & 46.68 & 65.67 & 44.34 \\
     &  & \ding{51} &  & \ding{51} &  & \cellcolor{second}46.75 & 65.55 & 45.02 \\
     &  & \ding{51} &  &  & \ding{51} & \cellcolor{best}\textbf{48.18} & \cellcolor{best}\textbf{67.27} & \cellcolor{best}\textbf{47.62} \\
    \hline
    \end{tabular}}
    \vspace{-10pt}
\end{table}

\noindent\textbf{Why choose ADSM and OT-MM?}

Beyond module-level ablations, we also compare ADSM and OT-MM with several representative alternatives, including standard Mamba scanning~\cite{gu2023mamba}, SS2D scanning~\cite{liu2024vmamba}, OT-LA~\cite{li2025alignmamba}, and vanilla Sinkhorn matching. Standard Mamba and SS2D mainly focus on generic sequence scanning, but they do not explicitly adapt the state update to modality-state changes, making them less effective when the sounding object switches over time. In contrast, ADSM explicitly modulates the effective step size according to current audio-visual cues and historical context, allowing the model to remain stable in steady segments while responding rapidly to abrupt state transitions. On the matching side, OT-LA and vanilla Sinkhorn can provide basic transport-based alignment, but they do not jointly model structural and temporal priors as in OT-MM. As a result, they are less effective in establishing robust instance-level correspondence across heterogeneous modalities. The results in Table~\ref{tab:scan_ot_comparison} verify our analysis, showing that the combination of ADSM and OT-MM consistently achieves the best performance.
\begin{figure}[t]
  \centering
  \includegraphics[width=\linewidth]{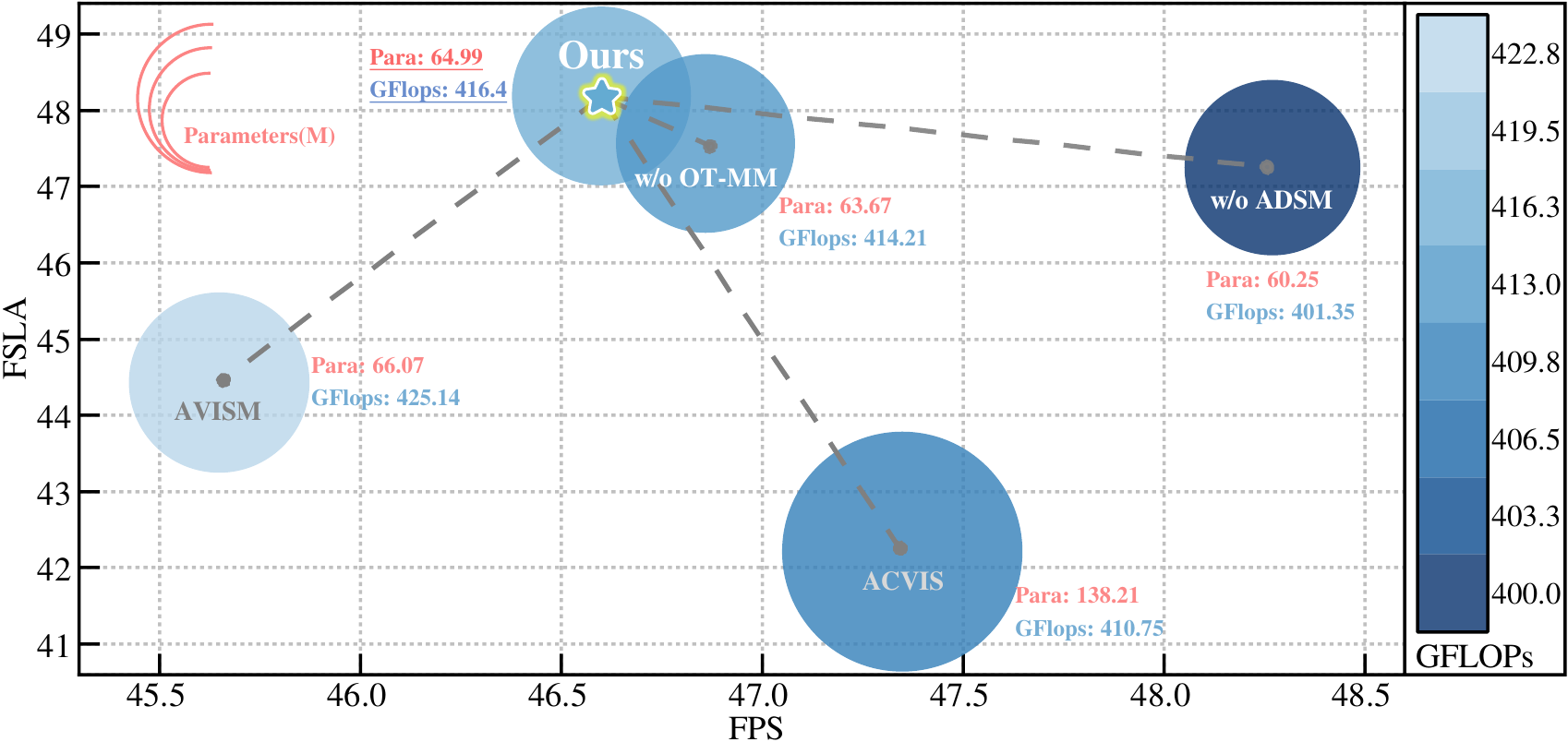}
  \vspace{-0.7cm}
  \caption{Performance-efficiency comparison of AVISM, ACVIS, Ours, and our variants w/o ADSM or OT-MM. Bubble size and color denote parameters and GFLOPs, respectively.}
\vspace{-10pt}
  \label{Pao}
\end{figure}

\begin{figure}[t]
  \centering
  \includegraphics[width=\linewidth]{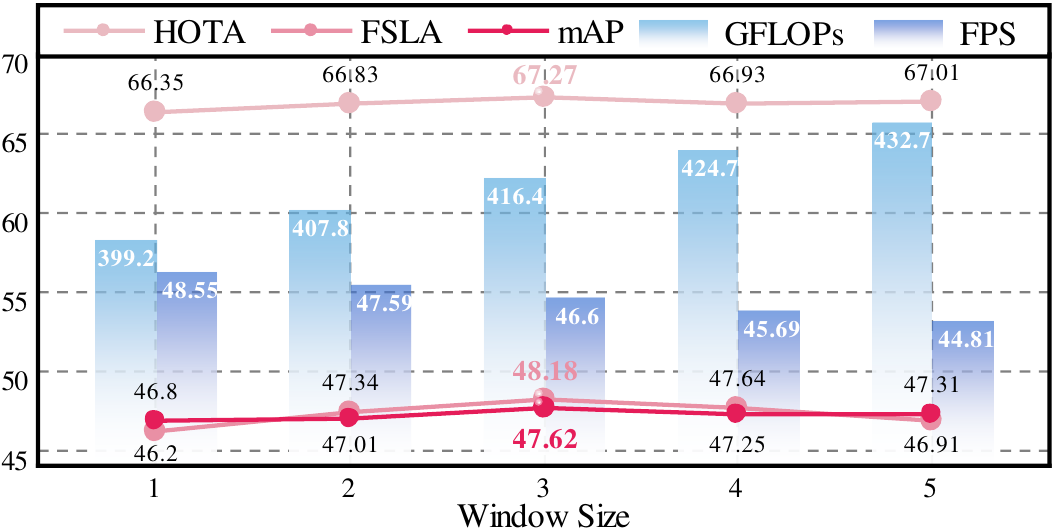}
  \vspace{-0.7cm}
  \caption{Effect of window size on performance and efficiency in the hierarchical temporal tracker. $w=3$ achieves the best trade-off among FSLA, HOTA, mAP, GFLOPs, and FPS.}
\vspace{-10pt}
  \label{ws}
\end{figure}

\noindent\textbf{Internal ablation of ADSM.}
To examine the effectiveness of ADSM, we analyze its two key components: the state perception branch (SP), implemented by the joint use of $s_{\text{modality}}$ and $s_{\text{diff}}$, and the Delta Gating Router (DGR). As shown in Table~\ref{tab:ablation_study_adsm}, enabling either SP or DGR alone already brings measurable gains over the ADSM baseline, while combining both yields the strongest performance. These results indicate that accurate perception of modality-state changes and adaptive fusion of current and historical cues are both essential for effective step-size modulation. 

\noindent\textbf{Why use Max norm and Cosine distance?}

Furthermore, we assess the state perception design in ADSM under alternative distance formulations. For $s_{\text{modality}}$, we compare the selected Max norm with L1 and L2 norms. The Max norm is adopted because abrupt modality-state transitions are better characterized by the dominant change across adjacent steps, whereas L1 and L2 tend to average out such localized but decisive variations. For $s_{\text{diff}}$, we compare the selected Cosine formulation with Euclidean and Pearson alternatives. Cosine distance captures semantic inconsistency between the audio-conditioned and raw visual representations while being less sensitive to magnitude variation. As shown in Fig.~\ref{cycle}, the selected formulations achieve the best overall performance, confirming their effectiveness for modeling modality-state changes and cross-modal discrepancy in AVIS.

\noindent\textbf{Internal ablation of OT-MM.}
To investigate which components are critical in OT-MM, we analyze the relative positional prior (RPP) and temporal consistency prior (TCP). RPP constrains structurally plausible audio-visual matching within each frame, while TCP improves temporal stability by encouraging consistent correspondence across adjacent time steps. As shown in Table~\ref{tab:ablation_study_otmm}, introducing either RPP or TCP improves performance over the OT-MM baseline, and combining both yields the best results. These results confirm that the two priors play complementary roles in establishing reliable transport-based instance-level matching over long video sequences. \emph{Additional ablations of OT and the MMD regularizer contributions are provided in the supplementary material.}

\subsection{Efficiency Analysis}
\noindent\textbf{Can the performance gains be obtained efficiently?} 
Beyond accuracy, we analyze the computational cost of our framework by comparing it with AVISM, ACVIS, and our variants without ADSM or OT-MM. As shown in Fig.~\ref{Pao}, our method achieves a more favorable performance-efficiency trade-off than the existing AVIS baselines, delivering stronger segmentation and tracking performance with competitive parameter count, FLOPs, and inference speed. Moreover, the module-level comparison shows that both ADSM and OT-MM introduce limited overhead while bringing clear gains in FSLA, HOTA, and mAP. These results indicate that the proposed modules improve performance in a cost-effective manner rather than through excessive computational complexity.

\begin{figure}[t]
  \centering
  \includegraphics[width=\linewidth]{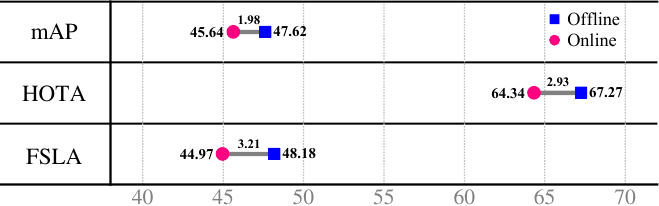}
  \vspace{-0.7cm}
  \caption{Comparison between offline and online settings. }
\vspace{-20pt}
  \label{on}
\end{figure}
\section{Discussion and Future Work}

Although our framework is developed in the offline AVIS setting, we observe that it retains reasonable performance after removing future-frame dependency and switching to causal inference. As shown in Fig.~\ref{on}, the online variant remains competitive, although a noticeable gap to the offline version persists. This suggests that state-aware temporal modulation and explicit cross-modal alignment may also be beneficial for online or streaming audio-visual understanding, while more effective causal memory propagation and incremental matching remain to be explored. In addition, the current hierarchical temporal tracker adopts a fixed local window size. As shown in Fig.~\ref{ws}, $w=3$ achieves the best trade-off in our experiments. But the optimal temporal receptive field may vary with the current modality state: rapidly changing segments may favor shorter windows, whereas stable segments may benefit from longer ones. We believe that extending the framework toward state-aware dynamic window selection is a promising future direction.

\section{Conclusion}
We propose a novel framework for audio-visual instance segmentation (AVIS) that addresses two core challenges: complex modality-state changes in long-range modeling and structural and distributional discrepancies across heterogeneous modalities. Our framework integrates Adaptive Dynamic Step Modulation (ADSM) for state-aware temporal step-size modulation and Optimal Transport-based Matching Modulation (OT-MM) for explicit instance-level audio-visual consistency with MMD regularization. Extensive experiments demonstrate that our method achieves state-of-the-art performance, while qualitative analyses further verify its effectiveness, especially in long-form and multi-instance scenarios.

\begin{acks}
This work was supported in part by the National Natural Science Foundation of China under Grants 62372080 and 62376050, and in part by the Natural Science Foundation of Liaoning Province under Grant 2024-MSBA-24.
\end{acks}

\bibliographystyle{ACM-Reference-Format}
\bibliography{samot}


\end{document}